\documentclass[prl, twocolumn,amsmath,amssymb]{revtex4-1}
\usepackage{graphicx, color}
\usepackage{dcolumn}
\usepackage{bm}
\usepackage{epstopdf}
\usepackage{xcolor}
\usepackage{units}
\usepackage{subfigure}
\definecolor{TTH-color}{named}{green}
\definecolor{PV-color}{rgb}{0.97,0.57,0.11}
\definecolor{YL-color}{RGB}{128,128,0}
\definecolor{TTH-color2}{named}{red}
\definecolor{PV-color2}{rgb}{0.87,0.47,0.01}
\definecolor{YL-color2}{RGB}{128,128,0}

\renewcommand{\Re}{\mathop{\mathrm{Re}}}

\begin{document}
\title{Directly probing the chirality of Majorana edge states} 
\author{Yao Lu}
\affiliation{Department of Physics and Nanoscience Center, University of Jyväskylä, P.O. Box 35 (YFL), FI-40014 University of Jyvaskyla, Finland}

\author{P. Virtanen}
\affiliation{Department of Physics and Nanoscience Center, University of Jyväskylä, P.O. Box 35 (YFL), FI-40014 University of Jyvaskyla, Finland}

\author{Tero T. Heikkil\"a}
\affiliation{Department of Physics and Nanoscience Center, University of Jyväskylä, P.O. Box 35 (YFL), FI-40014 University of Jyvaskyla, Finland}

\date{\today}
\begin{abstract}
We propose to directly probe the chirality of  Majorana edge states in 2D topological superconductors using polarization selective photon absorption. When shining circularly polarized light on a 2D topological superconductor in disk geometry, the photons can excite quasiparticles only when the polarization of the light  matches the chirality of the Majorana edge states 
required by the angular momentum conservation. Hence one can obtain the chirality of the Majorana edge states by measuring the photon absorption rate. We show that the polarization selective photon absorption can also serve as a smoking gun evidence of the chiral Majorana edge mode. Our results pave a new way of detecting, investigating and manipulating the chiral Majorana edge states.

\end{abstract}

\maketitle

{\em Introduction}. Searching for Majorana fermions is an important topic in condensed matter physics because of its potential application in topological quantum computing based on the non-Abelian statistics \cite{kitaev2001unpaired,qi2011topological,fu2008superconducting,sau2010generic,law2009majorana,mourik2012signatures,nadj2014observation,sarma2015majorana,lu2016platform,manna2020signature,ezawa2020non,lian2018topological}. Recently, the chiral Majorana edge states (CMES) localized at the boundary of a 2D topological SC have attracted much attention \cite{he2017chiral,lu2020effect,qi2010chiral,wang2015chiral,kezilebieke2020topological,palacio2019atomic,kezilebieke2020}. However, its identification is challenging due to the neutrality of the zero energy Majorana state.

The probable CMES was first observed in the quantum anomalous Hall (QAH)/SC structure \cite{he2017chiral}. Tuning the external magnetic field, a half-quantized conductance plateau emerges as a signature of the CMES when the system is driven into a topological superconducting phase \cite{wang2015chiral}. However, it has been shown that there are also other possible origins of the half-quantized conductance plateau \cite{ji20181,huang2018disorder}, such as disorder.  Another platform realizing the CMES is the ferromagnet/Rashba SC bilayer structure \cite{yamakage2012evolution,kezilebieke2020topological,palacio2019atomic}. In this setup, using scanning tunneling microscopy (STM) and scanning tunneling spectroscopy (STS), finite density of states (DOS)  below the superconducting gap was observed only close to the edge of the ferromagnet and this is regarded as a signature of the CMES. Recently, it was proposed that the CMES can generate enhanced local optical response at finite frequencies, \cite{he2021optical,he2021local} as the states at nonzero energy can couple to the electromagnetic field. The local optical conductivity scales as the square of the frequency also distinguishing it from the normal edge states.

Despite the great progress made in investigating the properties of the CMES, the STM and local optical measurement only show finite DOS at low energies but fail to tell whether the low energy mode is a chiral mode propagating only in one direction. The main reason of the difficulty is the neutrality of the zero energy Majorana mode. In QAH materials, the chirality of the topological edge states can be probed by measuring the quantized Hall conductance \cite{chang2013experimental,liu2016quantum}. However, in a topological SC, the zero energy Majorana mode is charge neutral and hence produces no Hall conductance. Instead, the CMES generates a quantized thermal Hall conductance \cite{sumiyoshi2013quantum,shimizu2015quantum}, which is difficult to measure in experiments.

\begin{figure}[h!]
\centering
\subfigure{\includegraphics[width = 0.48\columnwidth]{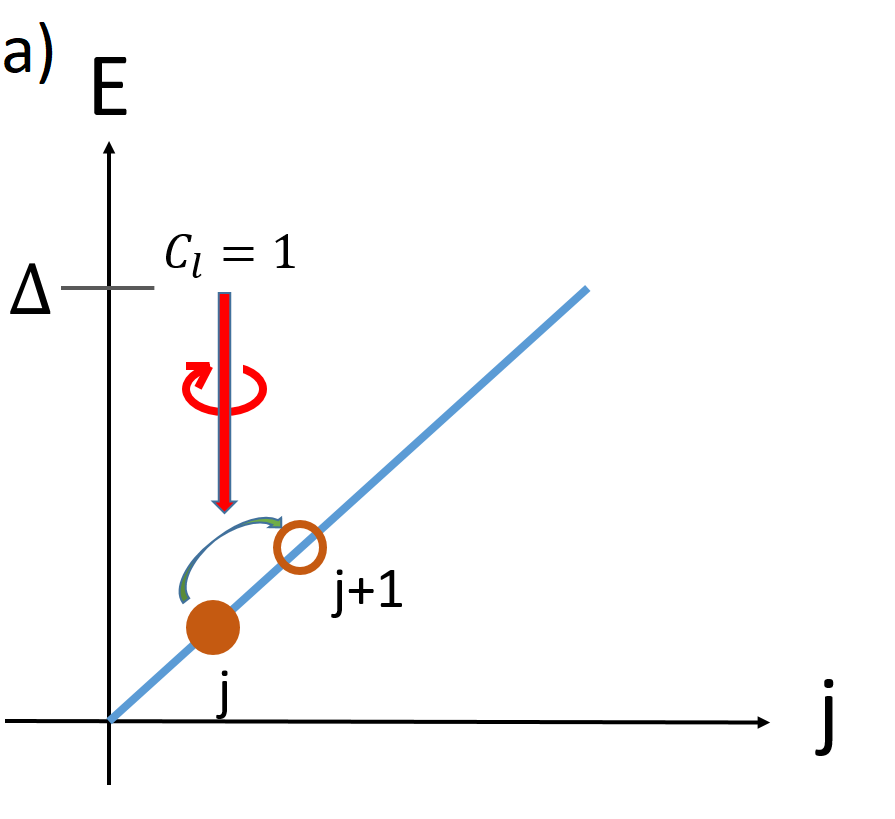}}
\subfigure{\includegraphics[width=0.48\columnwidth]{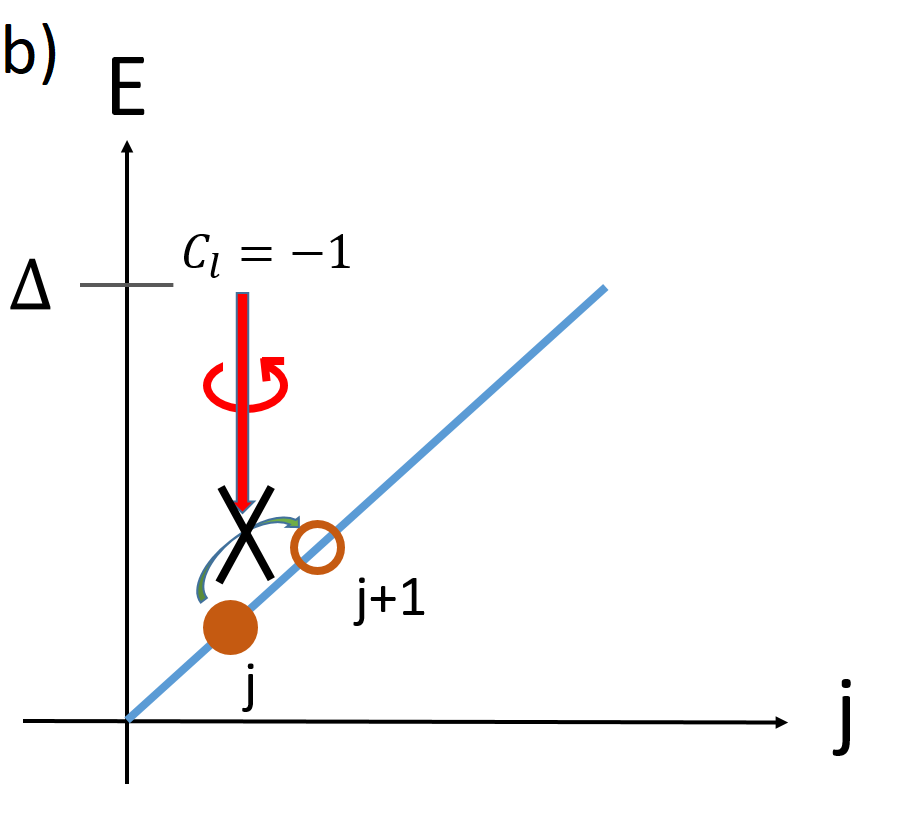}}
\caption{Schematic energy spectrum of the CMES. (a) A photon can excite a quasiparticle to a higher energy state if $C_l=C$. (b) The optical excitation is forbidden if $C_l\ne{}C$.}\label{Fig:sketch}
\end{figure}

In this Letter, we propose to directly probe the chirality of the CMES using polarization selective photon absorption (PSPA). We consider a topological SC in disk geometry. This setup preserves a rotation symmetry (RS). Thus the Majorana states are labeled by the total angular momentum $j$ rather than the momentum $k$ as in the usual cases \cite{phong2017majorana}. Near zero energy, the CMES always has a linear dispersion $E_j=Cv_Mj$, where $C=\pm 1$ is the Majorana chirality and $v_M>0$ is the Majorana angular speed. Shining  circularly polarized light with angular momentum $C_l=\pm 1$ and frequency $\omega$ on the sample, a quasiparticle with energy $E_j$ and angular momentum $j$ can absorb a photon and be excited to the state with energy $E_j+\hbar\omega$ and angular momentum $j+C_l$ if $C_l=C$ as shown in Fig.~\ref{Fig:sketch}. On the other hand, if $C_l\ne{}C$, this optical excitation is forbidden by angular momentum conservation. Thence by measuring the photon absorption, one can directly probe the chirality of the CMES. 
This effect can hence serve as a smoking gun evidence of the CMES. The advantage of this method is that it excludes the possibility of low energy trivial Andreev bound states that would have chirality independent response. Such exclusion cannot be achieved by STM or local optical methods. We also show that the temperature dependence of the photon absorption distinguishes the CMES from the normal chiral edge states.

{\em Model}. We first consider the simplest model of a 2D topological SC: a 2D spinless chiral $p$ wave SC described by the Hamiltonian \cite{qi2011topological}

\begin{equation}
H=\sum_{\boldsymbol{k}}\tilde{\Psi}_{\boldsymbol{k}}^{\dagger}\left[\left(\frac{\hbar^2\boldsymbol{k}^2}{2m}-\mu\right)\tau_3+\frac{\Delta}{k_F}\left(k_x\tau_1-Ck_y\tau_2\right)\right]\tilde{\Psi}_{\boldsymbol{k}}.
\label{eq:pwaveHamiltonian}
\end{equation}
Here $\tilde{\Psi}_{\boldsymbol{k}}=\left[\psi_{\boldsymbol{k}},\psi_{-{\boldsymbol{k}}}^{\dagger}\right]^{\text{T}}$, where $\psi_{\boldsymbol{k}}^{\dagger}$ is the electron creation operator which creates one electron with momentum ${\boldsymbol{k}}$. $\tau$ is the Pauli matrix acting on the particle-hole space. $C=\pm 1$ is the chirality of the chiral $p$ wave SC. $m$, $\mu$ and $\Delta$ are electron effective mass, chemical potential and pairing gap, respectively. We consider a sample in disk geometry \cite{stone2004edge} with the boundary $x^2+y^2=R_0^2$, where $x$ and $y$ are spatial coordinates and $R_0$ is the radius. The energy of the bulk states can be obtained by diagonalizing the Hamiltonian. The bulk spectrum close to $k\approx k_F$ is the usual Bogoliubov spectrum and has the gap $\Delta$. To find the subgap edge states, it is convenient to use the polar coordinates ($r$, $\phi$). This system preserves a generalized RS: $\left[H,J\right]=0$, where $J=-i\partial_{\phi}-\frac{1}{2}\tau_3$ is the generalized angular momentum operator around the $z$ axis. Thus the total angular momentum $j$ is a good quantum number and the edge states take the form

\begin{equation}
    \Psi_{M,j}=e^{ij\phi}\left[\begin{array}{c}
\psi_{+,M,j}(r)e^{i\phi/2} \\
\psi_{-,M,j}(r)e^{-i\phi/2}
\end{array}\right],
\end{equation}
where $j=n+1/2$ with $n\in Z$. Here We assume that the radius of the sample $R_0$ is much larger than the Majorana localization length and the Fermi wave length $2\pi/k_F$, where $k_F$ is the Fermi momentum and $v_F$ is the Fermi velocity. By using the hard wall boundary condition $\Psi_j(r=R_0)=[0,0]^{\text{T}}$, we obtain the energy and the wavefunction of the low-energy edge states

\begin{equation}
E_j=\frac{Cj\Delta}{k_FR_0},\label{eq:energy}
\end{equation}

\begin{eqnarray}
    i\psi_{+,M,j}=\psi_{-,M,j}=\frac{1}{\sqrt{2}}[e^{\kappa_+ (r-R_0)}-e^{\kappa_- (r-R_0)}],\label{eq:wavefunction}
\end{eqnarray}
where $ \kappa_{\pm}=\frac{m}{\hbar^2}\left[\frac{\Delta}{k_F}\pm\sqrt{\frac{\Delta^2}{k_F^2}-\frac{2\hbar^2\mu}{m}}\right]$. One can see that there are low energy states localized at the boundary as long as $\mu>0$ required by $\text{Re}(\kappa_{\pm})>0$.  The effective Majorana angular speed is given by $v_M=\frac{C\Delta}{k_FR_0}$. This shows that the chirality of the CMES is indeed $C$.

{\em Optical response}. When shining light with a frequency $\omega$ much lower than the pairing gap of the sample, the photon can be absorbed by the edge states and excite the quasiparticles to higher-energy states, if selection rules allow it. For circularly polarized light, the vector potential is $\boldsymbol{A}=A(1,iC_l)$, where $C_l=\pm 1$ is the angular momentum (polarization) of the photon. The photon absorption rate $W$ can be determined by $W(\omega)=4\pi A^2\omega^2R_0^2\text{Re}\left[\sigma_{ll}(\omega)\right]$, where $\sigma_{ll}$ is the optical conductance, which is given by

\begin{eqnarray}
    \sigma(\omega)=\frac{i}{2\pi^2\omega R_0^2}\sum_{m,n}\langle m|J_l^{\dagger}|n\rangle\langle n|J_l|m\rangle \frac{f(E_m)-f(E_n)}{E_m-E_n-\hbar\omega+i0^+}, \label{eq:conductivity}
\end{eqnarray}
where $|m\rangle$ is the edge eigenstate with angular momentum $m$, $E_m$ is the eigenenergy and $f(E)$ is the Fermi distribution function.  The generalized current operator $J_l$ is the integral of the current density operator over the volume given by 

\begin{equation}
    J_l=\frac{\hbar e}{m} e^{iC_l\phi}\left(-i\partial_r+\frac{C_l}{R_0}\partial_{\phi} \right)\label{eq:current}.
\end{equation}
Here $e$ is the electron charge. With this generalized current operator we can calculate the matrix element in Eq.~(\ref{eq:conductivity}),
\begin{equation}
  \langle m|J_l|n\rangle=\frac{i\hbar ev_FE_n}{\Delta}\delta_{m,n+C_l},\label{eq:matrix}
\end{equation}
where $v_F=\hbar k_F/m$ is the Fermi velocity in the normal state. This  matrix element has several important features: first, it is proportional to $v_F$ rather than the velocity of the edge states; second, the matrix element is linear in energy $E_n$ in contrast to the case of normal chiral edge states \cite{he2021local}; third, the finiteness of the matrix element depends on the polarization of the light $C_l$. We get the optical conductance by substituting Eq.~(\ref{eq:matrix}) into Eq.~(\ref{eq:conductivity}),

\begin{eqnarray}
     \text{Re}\left[\sigma_{ll}(\omega)\right] \approx\frac{\pi^2e^2T^2v_F^2k_F}{3h\Delta^3  R_0}\delta\left(\hbar\omega-\frac{\Delta}{k_FR_0}\right)\delta_{C,C_l},\label{eq:conductivity2}
\end{eqnarray}
 where $f$ is the Fermi distribution function, and we assumed $\hbar\omega\ll{}k_BT$. $\text{Re}\left[\sigma_{ll}(\omega)\right]$ is nonzero only when $C=C_l$ and $\hbar\omega=\Delta/k_FR_0$ required by the energy conservation and angular momentum conservation. Therefore, shining circularly polarized light on a 2D topological superconductor one can directly probe the chirality of the CMES via measuring the photon absorption rate. 

In systems with higher Chern number \cite{kezilebieke2020}, several Majorana edge modes are present. If the velocities of the modes differ, they appear in the photon absorption rate as separate peaks, at the frequencies determined by the angular momentum selection rule. If the velocities are too close to each other to be resolved, the low-temperature absorption spectrum provides a lower bound for the number of edge states.

 For comparison with  Eq.~\eqref{eq:conductivity2}, we consider the optical response of normal chiral edge states in a QAH insulator. Note that a QAH insulator can also be described by Eq.~(\ref{eq:pwaveHamiltonian}) with $\tilde{\Psi}$ replaced by  $\Psi=(\psi_{\uparrow},\psi_{\downarrow})^{\text{T}}$, the Pauli matrix $\tau$ replaced by the Pauli matrix $\sigma$ acting on the spin space and $\Delta$ replaced by the spin orbit coupling strength $\Delta_{\alpha}$ \cite{qi2006topological}. The current operator however is different, and the optical conductance is given by \cite{he2021local}
\begin{equation}
    \text{Re}\left[\sigma_{ll}(\omega)\right]=\frac{ e^2\Delta}{hk_FR_0}\delta(\hbar\omega-\frac{\Delta}{k_FR_0})
\end{equation}
In the normal state, $\text{Re}\left[\sigma_{ll}\right]$ is independent of the temperature. Hence the temperature dependence of the optical response can be used to distinguish the CMES from the normal chiral edge states in a QAH insulator.

\begin{figure}
\centering
\subfigure{\label{a}\includegraphics[width = 1\columnwidth]{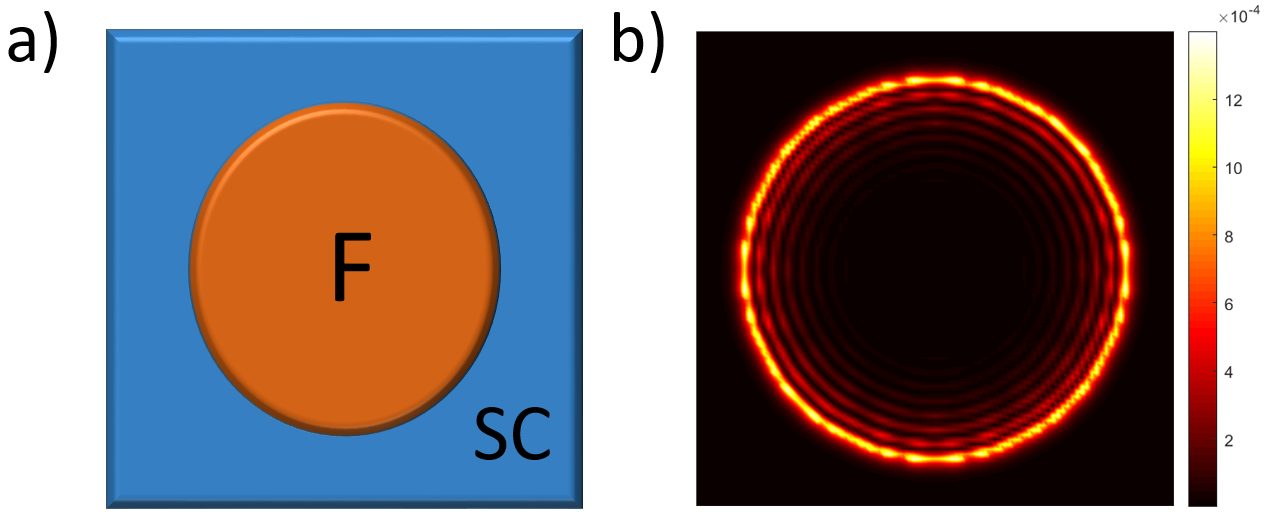}}
\caption{(a)Sketch of F/SC structure realizing 2D topological superconductor. (b) Spatial distribution of $|\Psi|^2$, where $\Psi$ is the wavefunction of the lowest energy state.}\label{Fig:FISC}
\end{figure}

\begin{figure}
\centering
\subfigure{\label{a}\includegraphics[width = 1\columnwidth]{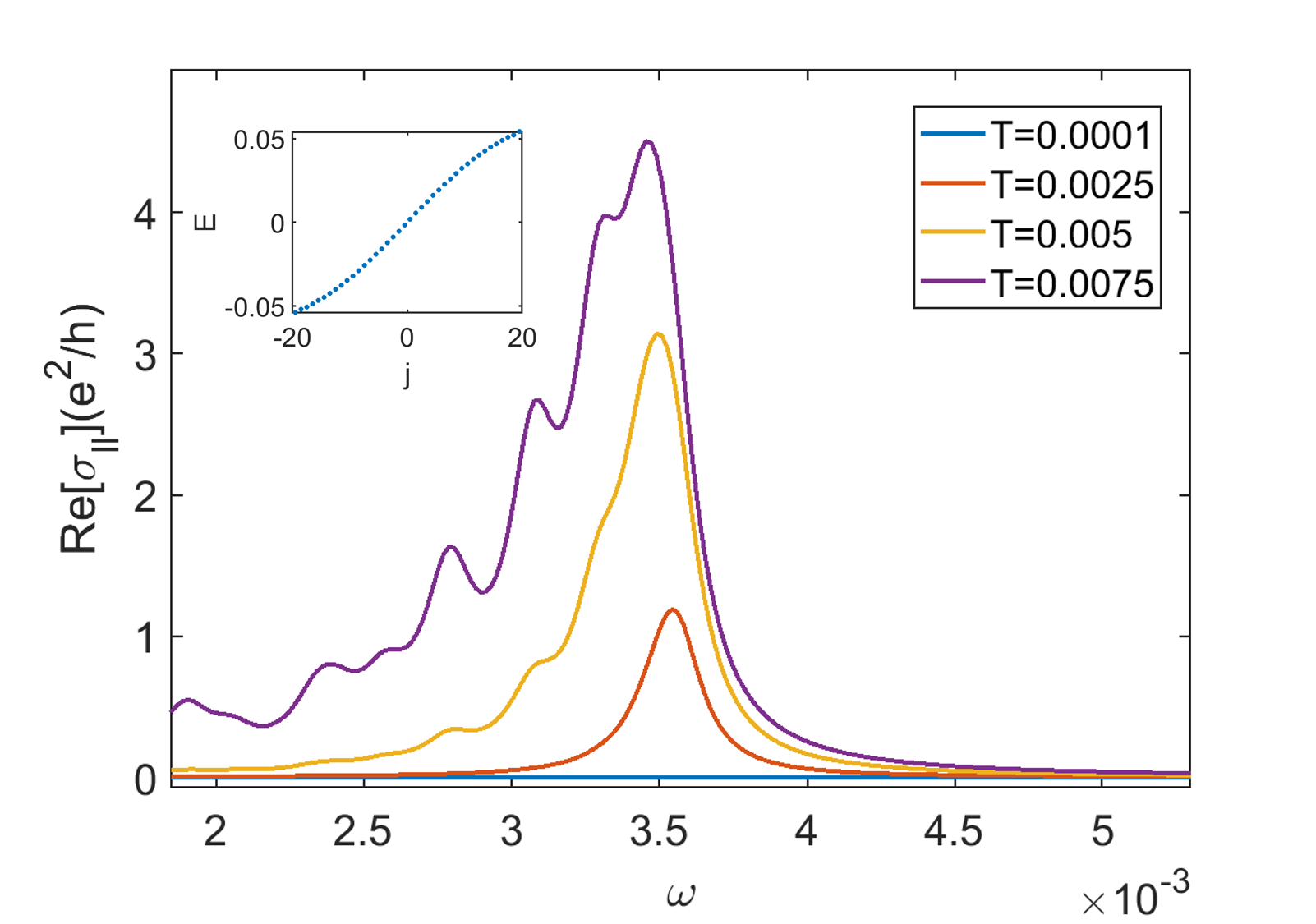}}
\caption{The real part of the optical conductance as a function of frequency at different temperatures calculated from the lattice model Eq.~(\ref{eq:TBM}). This figure only shows the optical conductance when $C_l=1$. For $C_l=-1$, $\text{Re}\left[\sigma_{ll}\right]$ overlaps the $C_l=1$, $T=0.0025$ curve. The inset shows the non-linear spectrum of the CMES. Parameters used here are: $t=1$, $\Delta=0.2$, $\mu=0$, $V_0=0.5$, $\alpha=0.25$, $R_0=40$.}\label{fig:conductivityT}
\end{figure}

{\em F/SC lattice model}. Recently there have been several experiments realizing the CMES in a ferromagnet/SC structure as shown in Fig.~\ref{Fig:FISC}a \cite{kezilebieke2020,palacio2019atomic}. This system can be described by the effective tight-binding Hamiltonian

\begin{eqnarray}
    H_{TB}&=&\sum_{\boldsymbol{R},\boldsymbol{d}}t\psi_{\boldsymbol{R}+\boldsymbol{d}}^{\dagger}\Psi_{\boldsymbol{R}}-\mu\Psi_{\boldsymbol{R}}^{\dagger}\Psi_{\boldsymbol{R}}-V_z(\boldsymbol{R})\Psi_{\boldsymbol{R}}^{\dagger}\sigma_z\Psi_{\boldsymbol{R}}\nonumber\\
    &+&i\alpha\boldsymbol{d}\times\hat{z}\cdot\Psi_{\boldsymbol{R}+\boldsymbol{d}}\boldsymbol{\sigma}\Psi_{\boldsymbol{R}}+\Delta\Psi_{\boldsymbol{R},\uparrow}\Psi_{\boldsymbol{R},\downarrow}+h.c.,\label{eq:TBM}
\end{eqnarray}
where $\boldsymbol{R}$ denotes the lattice sites, $\boldsymbol{d}$ denotes the two unit vectors $d_x$ and $d_y$. $t$ is the hopping strength, $\mu$ is the chemical potential, $\alpha$ is the Rashba coefficient  and $\Delta$ is the 
pair potential. $V_z(\boldsymbol{R})$ is position dependent exchange field $V_z(\boldsymbol{R})=V_0$ when $|\boldsymbol{R}|<R_0$, and otherwise zero. When $V_0^2>\mu^2+\Delta^2$, the ferromagnetic part is topological while the non-magnetic part is trivial.  We get the eigenstates of this tight-binding Hamiltonian by diagonalizing it. The wavefunction of the eigenstate closest to the zero energy is shown in Fig.~\ref{Fig:FISC}b. This subgap state localized at the boundary of the ferromagnet region is the chiral Majorana state. In this tight-binding model the generalized current operator corresponding to circularly polarized light is given by

\begin{eqnarray}
 J_l&=&\frac{ea}{4\pi }\sum_{\boldsymbol{R},q}B_q\left[it\Psi_{\boldsymbol{R}+\boldsymbol{d}_{q}}^{\dagger}\Psi_{\boldsymbol{R}}+h.c.\right]\nonumber\\
 &-&B_q\left[\alpha\boldsymbol{d}_q\times\hat{z}\cdot\Psi_{\boldsymbol{R}+\boldsymbol{d}_{q}}^{\dagger}\boldsymbol{\sigma}\Psi_{\boldsymbol{R}}+h.c.\right]\label{eq:current2}
\end{eqnarray}
where  $q=x,y$ and $B_x=1$, $B_y=i$. $a$ is the lattice constant. The first term in Eq.~(\ref{eq:current2}) comes from the kinetic energy and the second term is the Rashba contribution. We numerically calculate the optical conductance using Eq.~(\ref{eq:conductivity}) and the results are shown in Fig.~\ref{fig:conductivityT}. When $C_l=1$, $\text{Re}\left[\sigma_{ll}\right]$ is finite and has a peak when the photon frequency matches the energy spacing of the CMES. When $C_l=-1$, $\text{Re}\left[\sigma_{ll}\right]$ is close to zero. Increasing the temperature the conductance peak is enhanced and broadened. This broadness is due to the fact that the energy spacing of the CMES in this lattice model is not constant, as seen in the inset of Fig.~\ref{fig:conductivityT}.

\begin{figure}
\centering
\subfigure{\label{a}\includegraphics[width = 1\columnwidth]{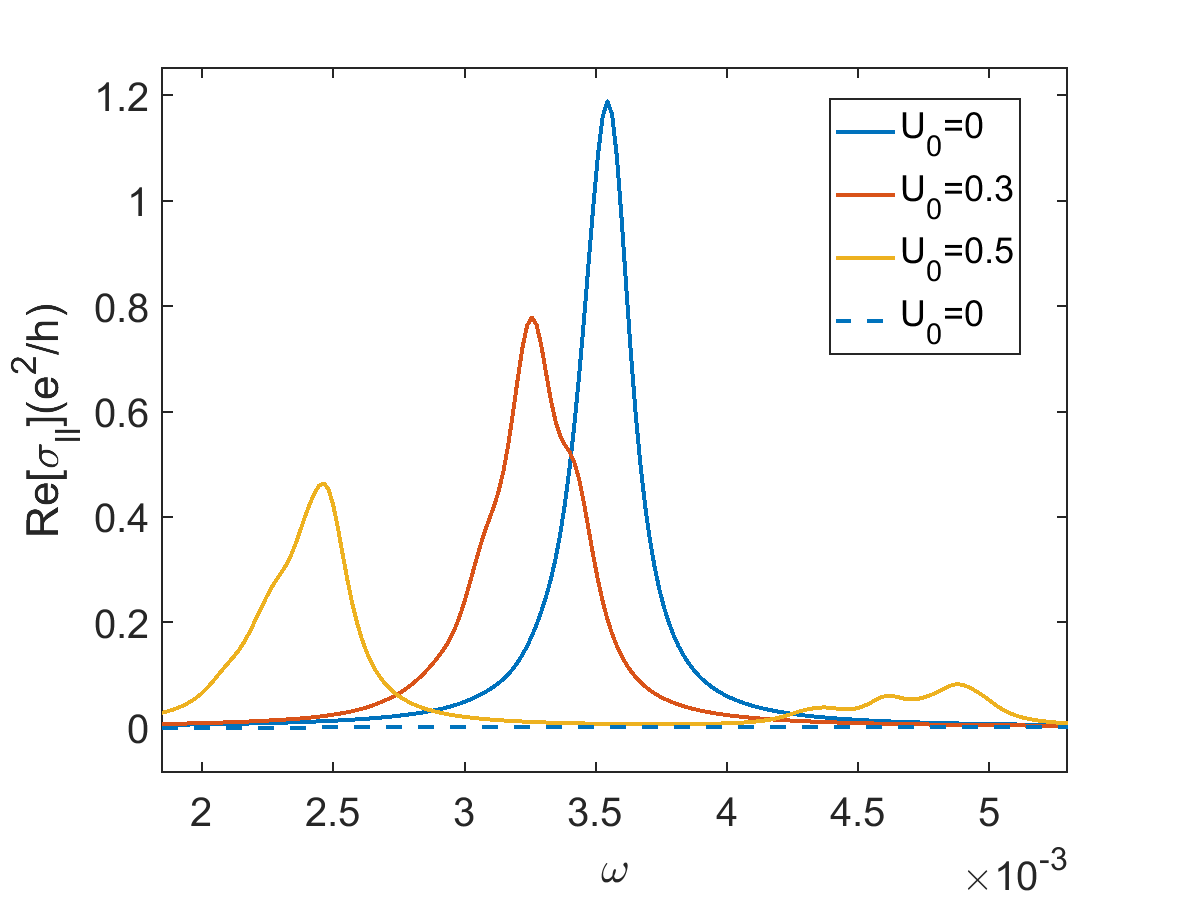}}
\caption{The real part of the optical conductance as a function of frequency for different disorder strengths. Here we consider one disorder configuration. The solid lines and dashed lines represent $C_l=1$ and $C_l=-1$, respectively. For $C_l=-1$, $\text{Re}\left[\sigma_{ll}\right]$ with any disorder strength is much smaller than that of $C_l=1$ and overlap each other, so we only show one of them in this figure. The parameters used here are the same as in Fig.~\ref{fig:conductivityT} with $T=0.0025$.}\label{fig:conductivitydis}
\end{figure}

{\em RS breaking}. So far we have considered the systems preserving the RS. To investigate how robust the PSPA is against the RS breaking, we first consider a case with non-magnetic impurities which locally break the RS but respect the RS on average. We consider the following Hamiltonian

\begin{figure}
\centering
\subfigure{\label{a}\includegraphics[width = 1\columnwidth]{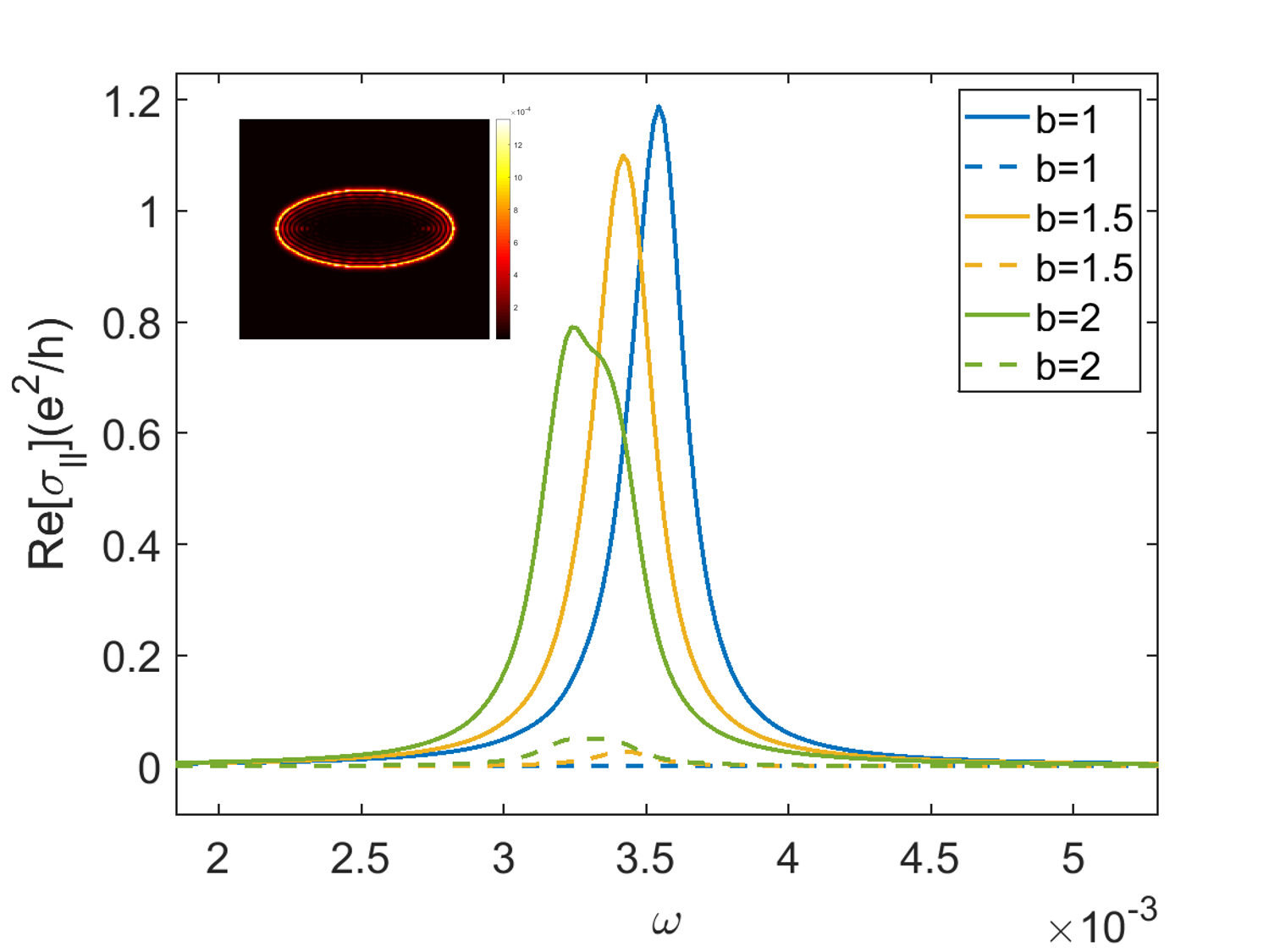}}
\caption{Real part of the optical conductance rate as a function of frequency for different shapes. The solid lines and dashed lines represent $C_l=1$ and $C_l=-1$, respectively. The inset shows the spatial distribution of the lowest energy for $b=1$. The parameters used here are the same as in Fig.~\ref{fig:conductivityT} with $T=0.0025$. }\label{fig:conductivityshape}
\end{figure}

\begin{equation}
    H_{TBD}=H_{TB}-\sum_{\boldsymbol{R}}U(\boldsymbol{R})\Psi_{\boldsymbol{R}}^{\dagger}\Psi_{\boldsymbol{R}},
\end{equation}
where $U(\boldsymbol{R})$ is the disorder potential which is uniformly distributed in $[-U_0,U_0]$. We numerically calculate $\text{Re}\left[\sigma_{ll}\right]$ in the presence of $U(\boldsymbol{R})$ and the results are shown in Fig.~\ref{fig:conductivitydis}. It shows that the effect of the disorder is to shift the position of the conductance peak and broaden it. The PSPA survives even when the disorder strength is larger than the pairing gap. This result is expected because the chiral edge states are topologically protected, and hence the PSPA is robust against disorder.

Next we consider an elliptical ferromagnetic island with a boundary described by $bR_x^2+\frac{1}{b}R_y^2=R_0^2$. When $b\neq 1$, this geometry does not respect the RS and the angular momentum is no longer a good quantum number. We numerically calculate $\text{Re}\left[\sigma_{ll}\right]$ and show the results in Fig.~\ref{fig:conductivityshape}. One can see that even with a large shape deformation ($b=2$) which greatly breaks the rotation symmetry (inset of Fig.~\ref{fig:conductivityshape}), $\text{Re}\left[\sigma_{ll}\right]$ for $C_l=1$ is much larger than that for $C_l=-1$. This can be understood as follows. Since the Majorana wave function is single valued, we can use an integer winding number to label the CMES. The circular polarized light with $C_l$ approximately changes the winding number by $C_l$ and we assume the CMES $\Psi_m$ with the winding number $w$ roughly satisfies the normalization condition $\langle\Psi_m|\Psi_{m'}\rangle\approx\delta_{m,m'}$. Therefore, similar to the angular momentum conservation, the "winding number conservation" can also induce approximatively PSPA.

\begin{figure}
\centering
\subfigure{\label{a}\includegraphics[width =1\columnwidth]{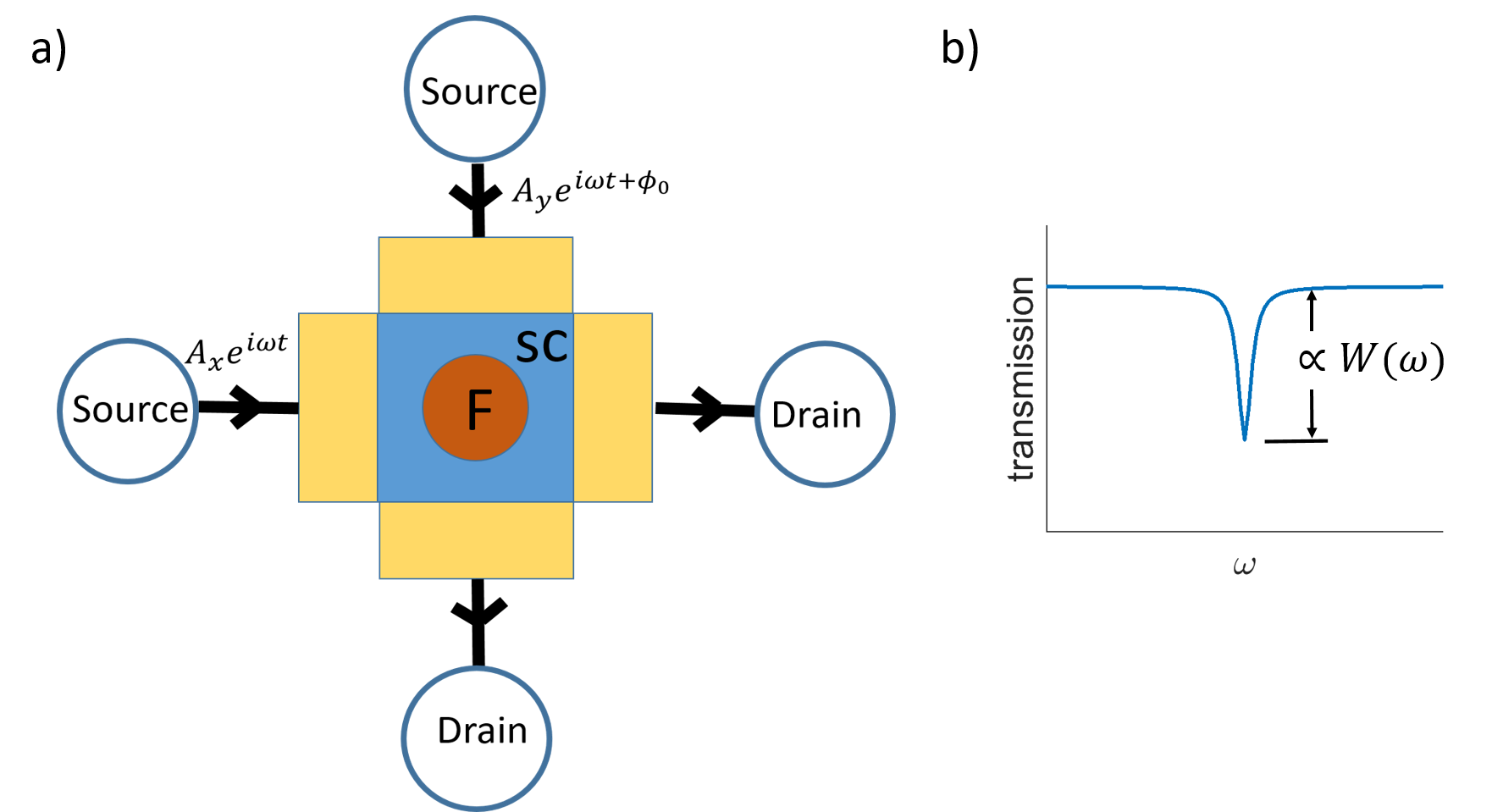}}
\caption{a) Setup to probe the chirality of the CMES. The yellow color denotes the electrodes. b) Sketch of predicted microwave transmission rate as a function of frequency $\omega$. There is a transmission dip when the frequency is equal to the energy spacing of the CMES.}\label{fig:setup}
\end{figure}

{\em Experimental detection}. Polarized electromagnetic fields can be generated in a cross-shaped rf cavity, \cite{alegre2007} as illustrated in Fig.~\ref{fig:setup}. The polarization is controlled by the phase shift $\phi_0$ between the sources, $A_x(t)\sim{}A\Re e^{i\omega t}$, $A_y(t)\sim{}A\Re e^{i\omega t+\phi_0}$, and is circularly polarized if $\phi_0=\pm \pi/2$. As we have shown, if the microwave polarization matches the chirality of the CMES, the photon absorption rate is peaked at frequency $\hbar\omega=\delta E$, corresponding to the energy spacing of the CMES, $\delta E\simeq{}\Delta/(\hbar k_F R_0)\sim\unit[10]{MHz}\text{--}\unit[1]{GHz}$ for $\Delta\sim\unit[0.1]{meV}$ \cite{kezilebieke2020} and island sizes $R_0\sim{}10^2\text{--}10^4\times{}k_F^{-1}$. This can be observed as an enhanced absorption peak, whose presence and amplitude depends on $\phi_0$. The resonant frequency can also be varied in situ, since the superconducting gap $\Delta$ depends on e.g. temperature $T$ and applied magnetic field $B$.

{\em Conclusion}. To conclude, we have shown that when shining circularly polarized light onto a 2D topological superconductor with subgap CMES, the photon absorption rate is finite only when the polarization of the light matches the chirality of the CMES. We propose that this effect can be used to directly probe the chirality of the CMES. Importantly, we show that this selective photon absorption effect is robust against rotation symmetry breaking, 
such as disorder or shape deformation.

\begin{acknowledgments}
We thank Shawulienu Kezilebieke, Mika Sillanpää and Juha Muhonen for discussions. This  work  was  supported  by the Academy of Finland (project number 317118). It has also received funding from the European Union’s Horizon 2020 research and innovation programme under grant agreement No. 800923.
\end{acknowledgments}

\bibliography{refs}

\section{Supplemental Material}
\subsection{Chiral Majorana states in $p$ wave superconductors}

Consider a 2D spinless chiral $p$ wave SC described by the Hamiltonian

\begin{equation}
H=\sum_{\boldsymbol{k}}\tilde{\Psi}_{\boldsymbol{k}}^{\dagger}\left[\left(\frac{\hbar^2\boldsymbol{k}^2}{2m}-\mu\right)\tau_3+\frac{\Delta}{k_F}\left(k_x\tau_1-Ck_y\tau_2\right)\right]\tilde{\Psi}_{\boldsymbol{k}}.
\label{eq:pwaveHamiltonian}
\end{equation}
Here $\tilde{\Psi}_{\boldsymbol{k}}=\left[\psi_{\boldsymbol{k}},\psi_{-{\boldsymbol{k}}}^{\dagger}\right]^{\text{T}}$, where $\psi_k^{\dagger}$ is the electron creation operator which creates one electron with momentum ${\boldsymbol{k}}$. $\tau$ is the Pauli matrix acting on the particle-hole space. $C=\pm 1$ is the chirality of the chiral $p$ wave SC. $m$, $\mu$ and $\Delta$ are electron effective mass, chemical potential and pairing gap, respectively. We consider a sample in disk geometry with the boundary $x^2+y^2=R_0^2$, where $x$ and $y$ are spatial coordinates and $R_0$ is the radius. The bulk states can be obtained by diagonalizing the Hamiltonian in ${\boldsymbol{k}}$ space. The energy spectrum of the bulk is given by

\begin{equation}
    E_{Bulk,\boldsymbol{k}}=\pm\sqrt{\Delta_p^2+\epsilon_k^2},
\end{equation}
where $\Delta_p=\Delta|\boldsymbol{k}|/k_F$ and $\epsilon_k=\frac{\boldsymbol{k}^2}{2m}-\mu$.  To find the subgap edge states, it is convenient to use the polar coordinates ($r$, $\phi$), \begin{equation}
k_x\rightarrow -i\cos(\phi)\partial_r+i\sin(\phi)\frac{1}{r}\partial_{\phi},
\end{equation}

\begin{equation}
k_y\rightarrow -i\sin(\phi)\partial_r-i\cos(\phi)\frac{1}{r}\partial_{\phi},
\end{equation}

and

\begin{equation}
    \boldsymbol{k}^2\rightarrow -\partial^2_r-\frac{1}{r}\partial_r-\frac{1}{r^2}\partial^2_{\phi}.
\end{equation}
The system under consideration preserves a generalized rotation symmetry 
\begin{equation}
    \left[H,J\right]=0,
\end{equation}
where $J$ is the angular momentum operator

\begin{equation}
    J=-i\partial_{\phi}-\frac{1}{2}\sigma_z.
\end{equation}
 Thus the eigen-states of $H$ must be eigen-states of $J$. We write the eign-states as

\begin{equation}
    \Psi_j=e^{ij\phi}\left[\begin{array}{c}
\psi_{+,j}e^{i\phi/2} \\
\psi_{-,j}e^{-i\phi/2}
\end{array}\right],
\end{equation}
where $j=n+1/2$ with $n\in Z$. Substituting the above expression back into the Schrodinger equation we have
\begin{widetext}
\begin{eqnarray}
 \left[\begin{array}{cc}
-\frac{\hbar^2}{2m}(\partial^2_r+\frac{1}{r}\partial_r-\frac{(j+1/2)^2}{r^2})-\mu & \frac{\Delta}{k_F}(-i\partial_r+C\frac{j-1/2}{r})e^{iC\phi}\\
\frac{\Delta}{k_F}(-i\partial_r-C\frac{j+1/2}{r})e^{-iC\phi} & \frac{\hbar^2}{2m}(\partial^2_r+\frac{1}{r}\partial_r-\frac{(j-1/2)^2}{r^2})+\mu
\end{array}\right]
\left[\begin{array}{c}
\psi_{+,j}e^{i\phi/2} \\
\psi_{-,j}e^{-i\phi/2}
\end{array}\right]
=
    E_l\left[\begin{array}{c}
\psi_{+,j}e^{i\phi/2} \\
\psi_{-,j}e^{-i\phi/2}
\end{array}\right].
\end{eqnarray}
\end{widetext}
We assume that the size of the sample is much larger than the Majorana localization length and the Fermi wave length, so we can replace $r$ by $R_0$ and treat $1/R_0$ as a perturbation.
 First we consider the 0th order term with $j=0$, $E_j=0$.
\begin{equation}
  \left[\begin{array}{cc}
-\frac{\hbar^2}{2m}\partial^2_r-\mu & \frac{\Delta}{k_F}(-i\partial_r)\\
\frac{\Delta}{k_F}(-i\partial_r) & \frac{\hbar^2}{2m}\partial^2_r+\mu
\end{array}\right] \left[\begin{array}{c}
\psi_{+,j} \\
\psi_{-,j} 
\end{array}\right]
=0. \label{eq:eigeneq}
\end{equation}
We also note that the Hamiltonian in Eq.~\ref{eq:eigeneq} has a chiral symmetry with the operator $P=\tau_2$ and it can be off-diagonalized in the basis of the eigen-states of $P$ by a unitary transformation

\begin{equation}
    UHU^{-1}=\left[\begin{array}{cc}
0 & H_1\\
H_2 & 0
\end{array}\right]
\end{equation}

with

\begin{equation}
    H_1=\frac{\hbar^2}{2m}\partial_r^2-\frac{\Delta}{k_F}\partial_r+\mu,
\end{equation}

\begin{equation}
    H_2=\frac{\hbar^2}{2m}\partial_r^2+\frac{\Delta}{k_F}\partial_r+\mu.
\end{equation}
To find the edge states, we use the following trial wave function

\begin{eqnarray}
    \psi_0=\left[e^{\kappa_+(r-R_0)}-e^{\kappa_-(r-R_0)}\right]\left[\begin{array}{c}
u \\
v
\end{array}\right].
\end{eqnarray}
Here the real parts of $\kappa_1$ and $\kappa_2$ are both positive $\text{Real}(\kappa)>0$. Putting this trial wave function back into the eigen-equation we have

\begin{equation}
  \left[\begin{array}{cc}
0 & \frac{\hbar^2}{2m}\kappa^2-\frac{\Delta}{k_F}\kappa+\mu\\
\frac{\hbar^2}{2m}\kappa^2+\frac{\Delta}{k_F}\kappa+\mu & 0
\end{array}\right] \left[\begin{array}{c}
u \\
v 
\end{array}\right]
=0.
\end{equation}
We see that only when $\mu>0$ there are two solutions with positive real parts and $\mu>0$ is the topological regime. Here we assume $\mu>0$. It is straightforward to find the solution for $\kappa$ and also the approximate zero energy eigen-state.

\begin{equation}
    \kappa_{\pm}=\frac{m}{\hbar^2}\left[\frac{\Delta}{k_F}\pm\sqrt{\frac{\Delta^2}{k_F^2}-2\frac{\hbar^2\mu}{m}}\right]
\end{equation}

\begin{eqnarray}
    \Psi_0=\left[e^{\kappa_1(r-R)}-e^{\kappa_2(r-R)}\right]e^{ij\phi}\left[\begin{array}{c}
ie^{i\phi/2}\\
e^{-i\phi/2}
\end{array}\right]/\sqrt{2} .\label{eq:eigenstatesS}
\end{eqnarray}

Next we consider the first order perturbation in $1/R_0$. Here we also assume $j$ is small compared to $R_0k_F$, where $a$ is the lattice constant. The perturbation of the Hamiltonian is given by

\begin{equation}
    \delta H=\left[\begin{array}{cc}
-\frac{\hbar^2}{2mR_0}\partial_r & \frac{C\Delta}{k_F}\frac{i(j+1/2)}{R_0}e^{i\phi}\\
-\frac{C\Delta}{k_F}\frac{i(j-1/2)}{R_0}e^{-i\phi} & \frac{\hbar^2}{2mR_0}\partial_r
\end{array}\right].
\end{equation}
Using the non-degenerate perturbation theory we find the energy dispersion 

\begin{equation}
    E_l=\frac{jC\Delta}{k_FR}.\label{eq:energyS}
\end{equation}
From this expression one can see that $C$ is indeed the chirality of the Majorana mode.
\subsection{Optical conductance of the Majorana edge states in $p$ wave superconductors}

Shining circularly polarized light onto a 2D chiral $p$ wave superconductor, the Fourier amplitude of the induced gauge field with frequency $\omega$ is given by

\begin{equation}
    \boldsymbol{A}(\omega)=A(\hat{x}+iC_l\hat{y}), \label{eq:gaugeS}
\end{equation}
where $C_l$ is the polarization of the light. The Hamiltonian coupled to the gauge field is given by

\begin{equation}
H=\sum_{\boldsymbol{k}}\tilde{\Psi}_{\boldsymbol{k}}^{\dagger}\left[\left(\frac{(\boldsymbol{k}+e\boldsymbol{A}\tau_3)^2}{2m}-\mu\right)\tau_3+\frac{\Delta}{k_F}\left(k_x\tau_1-Ck_y\tau_2\right)\right]\tilde{\Psi}_{\boldsymbol{k}},
\label{eq:pwaveHamiltonian}
\end{equation}
where $e$ is the electron charge. Thus we can obtain the current operator

\begin{eqnarray}
J_l&=&\frac{\partial H}{\partial A}\nonumber\\
&=&\frac{e}{2m}\left[k_x+iC_lk_y\right].
\end{eqnarray}
In the polar coordinates we have

\begin{equation}
    J_l=\frac{e}{m} e^{iC_l\phi}\left(-i\partial_r+\frac{C_l}{R_0}\partial_{\phi} \right)\label{eq:currentS}.
\end{equation}

In the linear response theory, the optical conductance is given by

\begin{equation}
    \sigma_{ll}(\omega)=\frac{i}{2\pi^2 R_0^2\omega}\sum_{m,n}\langle m|J_l^{\dagger}|n\rangle\langle n|J_l|m\rangle\frac{f(E_m)-f(E_n)}{E_m-E_n-\omega+i0^+}, \label{eq:conductivityS}
\end{equation}
where $|m\rangle$ is the edge eigenstate with angular momentum $m$, $E_m$ is the eigenenergy and $f(E)$ is the Fermi distribution function. Substituting Eq.~¨(\ref{eq:eigenstatesS}), (\ref{eq:energyS}), (\ref{eq:currentS}) into Eq.~(\ref{eq:conductivityS}), we obtain the real part of optical conductance

\begin{eqnarray}
    \text{Re}\left[\sigma_{ll}(\omega)\right]&=&\frac{1}{2\pi R_0^2\omega \hbar}\sum_j\left[\frac{eV_FE_j}{\Delta}\right]^2\left[f(E_{j+1})-f(E_j)\right] \nonumber\\
     &&\times\delta(\omega-\frac{\Delta}{k_FR_0})\delta_{C,C_l}\nonumber \\
    &\approx&\frac{\pi^2e^2T^2v_F^2k_F}{3h\Delta^3  R_0}\delta\left(\hbar\omega-\frac{\Delta}{k_FR_0}\right)\delta_{C,C_l}.
\end{eqnarray}
Here we only calculate the edge contribution and have assumed $T\ll\Delta$.

\subsection{Current operator in the tight binding model}

The tight binding Hamiltonian coupled to the gauge field is given by

\begin{eqnarray}
    H_{TB}&=&\sum_{\boldsymbol{R},\boldsymbol{d}}te^{ie\boldsymbol{A}\cdot \boldsymbol{d}}\psi_{\boldsymbol{R}+\boldsymbol{d}}^{\dagger}\Psi_{\boldsymbol{R}}-\mu\Psi_{\boldsymbol{R}}^{\dagger}\Psi_{\boldsymbol{R}}-V_z(\boldsymbol{R})\Psi_{\boldsymbol{R}}^{\dagger}\sigma_z\Psi_{\boldsymbol{R}}\nonumber\\
    &+&i\alpha e^{ie\boldsymbol{A}\cdot \boldsymbol{d}}\boldsymbol{d}\times\hat{z}\cdot\Psi_{\boldsymbol{R}+\boldsymbol{d}}\boldsymbol{\sigma}\Psi_{\boldsymbol{R}}+\Delta\Psi_{\boldsymbol{R},\uparrow}\Psi_{\boldsymbol{R},\downarrow}+h.c.\label{eq:TBMS}
\end{eqnarray}

where $\boldsymbol{R}$ denotes the lattice sites, $\boldsymbol{d}$ denotes the two unit vectors $d_x$ and $d_y$. $t$ is the hopping strength, $\mu$ is the chemical potential, $\alpha$ is the Rashba coefficient  and $\Delta$ is the pairing gap. $V_z(\boldsymbol{R})$ is position dependent exchange field. $\boldsymbol{A}$ is the vector potential as defined in Eq.~(\ref{eq:gaugeS}). Thus the current operator $J=\frac{\partial H}{\partial \boldsymbol{A}}$ is given by

\begin{eqnarray}
 J_l&=&\frac{ae}{4\pi }\sum_{\boldsymbol{R},q}B_q\left[it\Psi_{\boldsymbol{R}+\boldsymbol{d}_{q}}^{\dagger}\Psi_{\boldsymbol{R}}+h.c.\right]\nonumber\\
 &-&B_q\left[\alpha\boldsymbol{d}_q\times\hat{z}\cdot\Psi_{\boldsymbol{R}+\boldsymbol{d}_{q}}^{\dagger}\boldsymbol{\sigma}\Psi_{\boldsymbol{R}}+h.c.\right].\label{eq:current2}
\end{eqnarray}
where $a$ is the lattice constant. In the manuscript we use this current operator to calculate the optical conductance in the tight binding model.

\end{document}